\documentclass[twocolumn,showpacs,preprintnumbers,aps,pra,amsmath,amssymb,floatfix]{revtex4-1}

\usepackage{graphicx}% Include figure files
\usepackage{dcolumn}% Align table columns on decimal point
\usepackage{bm}% bold math
\usepackage{epsfig}
\usepackage{color}
\usepackage{natbib}
\usepackage[colorinlistoftodos]{todonotes}

\newcommand{\ket}[1]{\left|{#1}\right\rangle}

\begin{document}

\title{Coherent spin mixing dynamics in thermal $^{87}$Rb spin-1 and spin-2 gases}

\author{Xiaodong He$^1$, Bing Zhu$^1$, Xiaoke Li$^1$, Fudong Wang$^1$, Zhi-Fang Xu$^2$ and Dajun Wang$^1$}
\email{djwang@phy.cuhk.edu.hk}
\affiliation{
$^1$Department of Physics, The Chinese University of Hong Kong, Hong Kong, China \\
$^2$Department of Physics and Astronomy, University of Pittsburgh, Pittsburgh, Pennsylvania 15260, USA
}

\begin{abstract}

We study the non-equilibrium coherent spin mixing dynamics in ferromagnetic spin-1 and antiferromagnetic spin-2 thermal gases of ultracold $^{87}$Rb atoms. Long lasting spin population oscillations with magnetic field dependent resonances are observed in both cases. Our observations are well reproduced by Boltzmann equations of the Wigner distribution function. Compared to the equation of motion of spinor Bose-Einstein condensates, the only difference here is a factor of two increase in the spin-dependent interaction, which is confirmed directly in the spin-2 case by measuring the relation between the oscillation amplitude and the sample's density.

\end{abstract}

\maketitle

\section{Introduction}

Ultracold spinor quantum gas has been a subject of growing interest in recent years due to its internal degrees of freedom which connect it naturally to quantum magnetism and many other important physics problems. Extensive experimental and theoretical investigations have been carried out on both the spinor gas itself and its various applications, such as ground state phase diagram, domain formation, topological excitations, spin squeezing, magnetometry and spin mixing dynamics~\cite{Kawaguchi2012,Kurn2013}. Central to the understanding of these intriguing phenomena is the spin-dependent interaction and its competition with other relevant energy scales~\cite{Ho1998,Tetsuo1998,Law1998,Pu1999}. Non-equilibrium coherent spin mixing dynamics, manifested as coherent spin population oscillations, is a direct paradigm of the spin-dependent interaction and its interplay with Zeeman energies \cite{Zhang2005,Kronjager2005}. Such dynamics have been observed in spin-1~\cite{Chang2005,Kronjager2005,Widera2005} and spin-2 $^{87}$Rb~\cite{Kronjager2006}, and spin-1 $^{23}$Na Bose-Einstein condensates (BECs)~\cite{Black2007} early on.

In analyzing the coherent spin mixing dynamics in spinor BECs, nice agreements can typically be obtained under the single-mode approximation (SMA), in which the external spatial and the internal spin degrees of freedom are separated from each other~\cite{Law1998,Chang2005,Kronjager2006,Black2007}. It is thus natural to ask whether such dynamics can be observed with multi-spatial-mode spinor gases. Although collisions in these gases are typically considered random or incoherent in the spatial degrees of freedom, coherence among spin degrees of freedom can persist for a long time~\cite{Deutsch2010,*Buning2011}. Indeed, coherent spin mixing dynamics were observed recently in a thermal spin-1 $^{23}$Na gas~\cite{Pechkis2013} and a quantum degenerate Fermi gas of $^{40}$K with large spin~\cite{Krauser2014,Ulrich2014}. Results from these experiments were explained well by dynamics in the spin degrees of freedom only, indicating spin and spatial modes can still be largely decoupled under right conditions even without condensates.

In this work, we present experimental investigations of coherent spin mixing dynamics in ultracold thermal $^{87}$Rb spinor gases. Different from the antiferromagnetic spin-1 Na spinor gas~\cite{Stenger1998}, the spin-dependent interaction in spin-1 $^{87}$Rb is ferromagnetic and typically much smaller in magnitude~\cite{Klausen2001,Kempen2002}. Amazingly, we still observe robust and long-time coherent spin mixing dynamics driven by the 70 pK spin-dependent interaction in thermal samples with a typical temperature of 400 nK. In addition, we also observe clean coherent spin mixing dynamics in spin-2 $^{87}$Rb thermal gas, which has more spin mixing channels and thus richer dynamics~\cite{Koashi2000,Ciobanu2000}. In both cases, dependences of the spin dynamics on external magnetic fields are studied in detail and are explained well by a theoretical model developed under a single-spatial-mode approximation.

The rest of this paper is organized as follows. In section II, we derive the collisionless Boltzmann equation for modeling the spin dynamics of thermal spin-1 and spin-2 Bose gases. In section III, we describe the experimental setup and data taking procedure. In section IV, experiment results are presented and compared with numerical simulations based on the model. We then conclude the paper in section V.

\section{Theory}
While spinor BECs are well described by coupled Gross-Pitaevskii equations, pure thermal spinor gases can be dealt with the semiclassical Boltzmann transport equation with the Wigner function as the distribution function. Following this approach, several groups have predicted the existing of spin waves and spin mixing oscillations in spin-1 thermal gases~\cite{Endo2008, Natu2010}. These theoretical results were successfully applied to the spin-1 Na thermal spinor gas in reference~\cite{Pechkis2013}. It was found that under the right experimental conditions, the spin dynamics can be separated from the multi-spatial modes. Compared with spin-1 spinor BECs, one needs only to multiply a factor of two in the spin-dependent interaction coefficient to account for the fact that thermal atoms are distinguishable~\cite{Pechkis2013}. Here we follow the same formalism but generalize it to include both spin-1 and spin-2 cases.

In the second-quantization language, the Hamiltonian for spin-1 and spin-2 atoms of mass $m$ in an external magnetic field $B$ can be expressed jointly as
\begin{equation}\label{Hamiltonian}
\begin{split}
  H=&\int d\mathbf{r} \bigg[\hat{\psi}_{k}^{\dag}\left(-\frac{\hbar^{2}}{2m}\nabla^{2}+V(\mathbf{r})+q F_{z}^2\right)\hat{\psi}_{k} \\
   &+\frac{g_{0}}{2}\sum\limits_{kj}\hat{\psi}_{k}^{\dag}\hat{\psi}_{j}^{\dag}\hat{\psi}_{j}\hat{\psi}_{k}
   +\frac{g_{1}}{2}\sum\limits_{kjlm}\hat{\psi}_{k}^{\dag}\hat{\psi}_{l}^{\dag} \mathbf{F}_{kj} \cdot \mathbf{F}_{lm} \hat{\psi}_{m}\hat{\psi}_{j}\nonumber\\
   &+\frac{g_{2}}{2}\sum\limits_{kj}\frac{1}{5}(-)^{k-j} \hat{\psi}_k^{\dag}\hat{\psi}_{-k}^{\dag}\hat{\psi}_{-j}\hat{\psi}_{j}\bigg],
\end{split}
\end{equation}
where $q = \frac{(g \mu_B B)^2}{\Delta E_{hf}}$ is the quadratic Zeeman energy, which is positive for spin-1 and negative for spin-2 $^{87}$Rb atoms. Here $g$ is the hyperfine Land\'e g-factor, $\mu_B$ is the Bohr magneton, and $\Delta E_{hf}$ is the ground-state hyperfine splitting. The linear Zeeman energy is gauged out due to total spin conservation. $V(\mathbf{r})$ is the external potential, $\mathbf{F}$ is the vector spin operator and $\hat{\psi}_{k}$ ($\hat{\psi}_{k}^{\dag}$) is the bosonic field annihilation (creation) operator for spin projection $k$. $\hbar$ is the reduced Planck's constant.

The interaction coefficients $g_0$, $g_1$, and $g_2$ can be expressed in terms of the s-wave scattering lengths $a_0$, $a_2$, and $a_4$ in the total spin $0$, $2$ and $4$ channels of two colliding atoms. For the spin-1 case, the total spin can be 0 and 2, and only $g_0 = 4\pi\hbar^2(a_0+2a_2)/3m$ which represents the density-dependent interaction, and $g_1 = 4\pi\hbar^2(a_2-a_0)/3m$ which represents the spin-dependent interaction, are present. The interaction between spin-1 $^{87}$Rb is ferromagnetic with $g_1$ negative. For the spin-2 case, the total spin can be 0, 2 and 4. Besides $g_0 = 4\pi\hbar^2(4a_2+3a_4)/7m$ and $g_1 = 4\pi\hbar^2(a_4-a_2)/7m$, there is also a third term $g_2 = 4\pi\hbar^2(7a_0-10a_2+3a_4)/7m$, which represents the singlet-pairing interaction. Spin-2 $^{87}$Rb is believed to be antiferromagnetic with $g_1$ positive and $g_2$ negative but of much smaller magnitude~\cite{Schmaljohann2004}.

Following the derivation of spin-1 Boltzmann equation~\cite{Natu2010,Endo2008,Pechkis2013}, we introduce the Wigner distribution function which has matrix elements
\begin{equation}
\begin{split}
f_{kj}(\mathbf{r},\mathbf{p},t)= & \int d\mathbf{r}' e^{-i\mathbf{p}\cdot\mathbf{r}'/\hbar} \langle\hat{\psi}_{j}^{\dag}(\mathbf{r}-\mathbf{r}'/2, t) \\
&\hat{\psi}_{k}(\mathbf{r}+\mathbf{r}'/2, t) \rangle.
\end{split}
\end{equation}
From the Wigner function, we can obtain experimental observables such as the local density $n_{kj}(\mathbf{r},t)\equiv \int d\mathbf{p}\,f_{kj}(\mathbf{r},\mathbf{p},t)/(2\pi\hbar)^3$. Taking the Hatree-Fock approximation, we have the standard collisionless Boltzmann equation of the Wigner function
\begin{equation}
\begin{split}
&\frac{\partial}{\partial t}f_{kj}+\frac{\mathbf{p}}{m}\cdot\nabla_{\mathbf{r}}f_{kj} -\nabla_{\mathbf{r}}V\cdot\nabla_{\mathbf{p}}f_{kj}-\frac{1}{i\hbar}[U,f]_{kj} \\
&-\frac{1}{2}\{\nabla_{\mathbf{r}}U,\nabla_{\mathbf{p}}f\}_{kj}=0,
\end{split}
\label{boltzmannequation}
\end{equation}
where we have introduced matrix elements of the interaction potential $U$,
\begin{equation}
\begin{split}
U_{kj}(\mathbf{r})=&\left[q F_z^2+g_0{\rm Tr}(n) +g_0n\right]_{kj} \\
&+\left[g_1\sum\limits_{\mu}{\rm Tr}\left(F_{\mu}n\right)F_{\mu} +g_1\sum\limits_{\mu}F_{\mu}nF_{\mu}\right]_{kj} \\
&+ 2g_2\frac{(-)^{k-j}}{5} n_{-j,-k}.
\end{split}
\label{Mmatrix}
\end{equation}
Here $\mu$ = $x$, $y$ and $z$, and Tr is the trace operation. The factor of two in the last term is a result of equal Hartree and Fock term contributions. Note that this term vanishes for spin-1 case.

In the limit of strong trapping potential where the spatial motion is moderately faster than the spin dynamics characterized by interaction energies $g_{1,2}{\rm Tr}(n)$, the spatial dependent interactions can be averaged out. Thus we can decouple the spinor evolution $\sigma_{kj}(t)=\sqrt{\rho_k}e^{i\theta_k}\sqrt{\rho_j}e^{-i\theta_j}$ from the spatial and momentum distribution $w(\mathbf{r},\mathbf{p})$ and express the Wigner function matrix elements as $f_{kj}(\mathbf{r},\mathbf{p},t)=w(\mathbf{r},\mathbf{p})\sigma_{kj}(t)$, in analogous to the popular SMA in dealing with spin dynamics in BECs. Substituting this into Eq.~(\ref{boltzmannequation}) and integrating over position and momentum, we obtain the following equation of motion for the coherent spinor dynamics in a thermal gas
\begin{equation}
\frac{\partial}{\partial t}\sigma_{kj}=\frac{1}{i\hbar} \left[U^{\rm spin },\sigma\right]_{kj},
\label{equationofmotion}
\end{equation}
where the spin-dependent interaction potential
\begin{equation}
\begin{split}
U^{\rm spin}_{kj}(\mathbf{r})= & \bigg[q F_z^2 + g_1\bar{n}\sum\limits_{\mu}{\rm Tr}\left(F_{\mu}\sigma\right)F_{\mu} + \\
& g_1\bar{n}\sum\limits_{\mu}F_{\mu}\sigma F_{\mu} \bigg]_{kj} + 2g_2\bar{n}\frac{(-)^{k-j}}{5}\sigma_{-j,-k}.
\end{split}
\end{equation}Here, $\bar{n}=\int d\mathbf{r}[{\rm Tr}(n(\mathbf{r}))]^2/N$ with $N$ the total atom number.

%\textcolor{blue}{To compare the dynamics between thermal gas and BEC, from corresponding GP equations describing the spinor dynamics in single-spatial-mode BEC, and by introducing the density matrix $\varsigma_{ab}=\phi^*_a \phi_b$, one can obtain the equations of motion $i\hbar\partial_t\varsigma =[U_{BEC},\varsigma]$ for spin-1 and spin-2 condensate},
%\textcolor{blue}{where
%\begin{eqnarray}
%\label{equationofmotionBEC}
%[U_{\rm BEC}]_{ab}=\left[q F_z^2+g_1\bar{n}\sum\limits_{\mu}{\rm Tr}\left(F_{\mu}\sigma\right)F_{\mu} \right]_{ab} \nonumber\\
%+g_2\bar{n}(-)^{a-b}\sigma_{-b,-a}.\quad
%\end{eqnarray}}

For spin-1, $\sigma$ is represented by 3$\times$3 matrices. With the identity $\sum\limits_{\mu}F_{\mu}A F_{\mu} = TrA+A+\sum\limits_{\mu}{\rm Tr}\left(F_{\mu}A\right)F_{\mu}$ for any 3$\times$3 matrix $A$, it can be easily shown that the two $g_1$ terms have equal contributions in Eq.~(\ref{equationofmotion}). Thus the spin-dependent interaction can be summed as $2g_1\bar{n}\sum\limits_{\mu}{\rm Tr}\left(F_{\mu}\sigma\right)F_{\mu}$, which is a factor of two larger than that in spin-1 pure BECs~\cite{Pechkis2013}.

%\textcolor{blue}{By comparing Eq.~(6) with Eq.~(\ref{equationofmotionBEC}),  the $g_2$ term in thermal gas is doubled.}

In the case of spin-2, $\sigma$ is represented by 5$\times$5 matrices and the above identity is not true in general. However, if $\sigma$ is constructed from pure state spinor wavefunctions, which is the case in our experiment, contributions from the two $g_1$ terms are again the same and thus the factor of two still holds. Combined with the factor of two in the $g_2$ term, the overall spin-dependent interaction is also doubled compared with spin-2 pure BECs.

\section{Experiments}
%%%%%%%%%%%%%%%%%%%%%%%%%%%%%%%%%%%%%%%%%%%%%%%%%%%%%%%%%%%%%%%%%%%%%%%%%%
%(rewrite)
\begin{figure*}
\centering\includegraphics[width=0.75\linewidth]{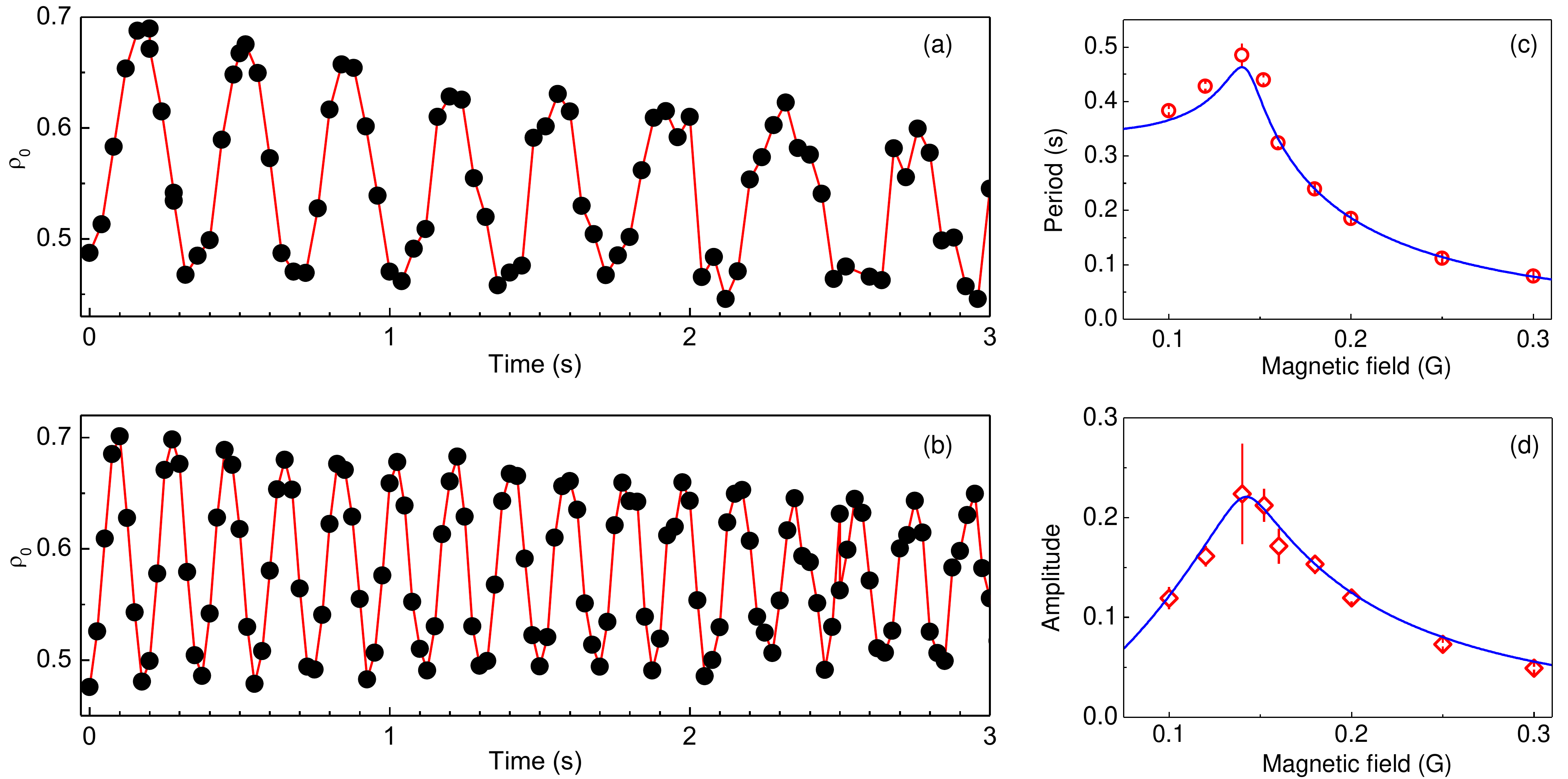}
\caption{(color online) Coherent spin population oscillation of $^{87}$Rb spin-1 normal gas and its dependence on magnetic field. (a) and (b) are exemplary temporal evolution of $\rho_0$ at magnetic fields of 0.1 G and 0.18 G, respectively. Red solid curves are for eye guiding. (c) and (d) show the magnetic field dependence of the oscillation period ($\circ$) and amplitude ($\diamond$) with a resonant feature located near 0.14 G. Error bars here are from fitting of the oscillations. Blue solid curves are fittings with~Eq.~(\ref{equationofmotion}).}
\label{fig1}
\end{figure*}

Our single vacuum chamber experimental setup has been descried before~\cite{Xiong2013a, Wang2013}. In brief, we prepare the ultracold $^{87}$Rb sample by evaporative cooling in a hybrid magnetic quadrapole plus optical dipole trap(ODT)~\cite{Lin2009,Xiong2013b,Wang2013} first. The magnetic trap ensures a hundred percent spin polarized sample in the $\ket{1,-1}$ hyperfine Zeeman state. The sample is then transferred to a crossed ODT in which the final evaporation is performed within a weak magnetic field to preserve the atom's spin state. In the same setup, we can produce a pure $^{87}$Rb BEC with $2\times10^5$ atoms. For the current experiment, we control the atom number and stop the evaporation before the BEC phase transition to make pure thermal gases. The measured typical final trap frequencies are $(\omega_x,\omega_y,\omega_z)=2\pi\times(190,211,113)\,\rm Hz$.

For investigating the spin-1 case, after the final evaporation, we hold the sample in the ODT for several hundred milliseconds to ensure full thermalization before the magnetic field is set to a desired value along the z-axis, with the transverse magnetic fields compensated to less than 3 mG. To initialize the spin dynamics, we directly apply a resonant radio frequency (rf) $\pi/2$-pulse to transfer the sample from $\ket{1,-1}$ hyperfine state to the fully transversely magnetized initial state $\zeta_1(0)=(1/2,1/\sqrt{2},1/2)^T$. Here $\zeta_1 = (\sqrt{\rho_{+1}}e^{-i\theta_{+1}},\sqrt{\rho_{0}}e^{-i\theta_{0}},\sqrt{\rho_{-1}}e^{-i\theta_{-1}})^T$ with $\rho_{-1,0,+1}$ and $\theta_{-1,0,+1}$ as fractional populations and phase in each spin components respectively. The system's magnetization is defined as $\rho_{+1}-\rho_{-1}$.

To investigate spinor dynamics in the spin-2 case, we first transfer the atoms to $\ket{2,-2}$ hyperfine state with a microwave rapid adiabatic passage at a $B$ field of 1.3 G with a near unity efficiency. The magnetic field is then changed adiabatically to a desired value before a fully transversely magnetized state $\zeta_2(0)=(1/4,1/2,\sqrt{3/8},1/2,1/4)^T$ is prepared with a rf $\pi/2$-pulse. The magnetization is defined as $2\rho_{+2}-2\rho_{-2}+\rho_{+1}-\rho_{-1}$ in this case.

After initialization, the system is in a far-from-equilibrium state and spin mixing dynamics will start. After holding the sample in the trap for a range of time for the dynamics to evolve, the ODT is turned off abruptly and atoms in different spin states are detected with the time-of-flight Stern-Gerlach absorption imaging technique after 12 ms expansion. Our absorption imaging setup is calibrated with the high saturation method~\cite{Reinaudi2007,Kwon2012}. The number of atoms in each spin state, $N_i$, is extracted from the images and the fractional population $\rho_i = N_i/N$ is then calculated.

\section{Results and discussions}

\subsection{Spin-1 case}

%%%%%%%%%%%%%%%%%%%%%%%%%%%%%%%%%%%%%%%%%%%%%%%%%%%%%%%%%%%%%%%%%%%%%%%

In Fig.~\ref{fig1}(a) and (b) we present spin population oscillations for $^{87}$Rb thermal spin-1 gases at magnetic fields of 0.1 G and 0.18 G, respectively. We can immediately see that the oscillation depends strongly on magnetic field. This can be understood from $U^{\rm spin}$ which has only the $2g_1$ term and the quadratic Zeeman $q$ term for spin-1. While the negative $2g_1$ term favors the ferromagnetic state, the positive $q$ favors the polar state~\cite{Ho1998}. Non-equilibrium spin mixing dynamics is just the result of the competition between them. Similar to spin-1 BEC, depending on their relative strengths, spin dynamics can be divided into the interaction regime and the Zeeman regime~\cite{Kronjager2005}. With the $2g_1$ term fixed by the density, external magnetic fields can be applied to tune the system into either regimes.

When the magnetic field is low and thus $q$ is small, the system is in the interaction regime and the oscillation period is predominately determined by the $2g_1$ term~\cite{Kronjager2005}. This is the case in Fig.~\ref{fig1}(a). At higher magnetic field B, $q$ ($\sim$72$B^2$ Hz/G$^2$) dominates the rather small $2g_1$ term ($2g_1\langle n\rangle$ $\sim$ 1.45 Hz for our typical density), and the system enters the Zeeman regime in which the oscillation period is $\propto 1/q$~\cite{Kronjager2005}. In both cases, these oscillations last for a rather long time, although only data in the first 3 seconds are shown. For longer time, the oscillation continues but becomes non-periodic. To extract the oscillation period and amplitude, we fit the first several oscillations to a damped sinusoidal function.

For the current initial state, the crossover between these two regimes happens when $q\approx2g_1\langle n\rangle$~\cite{Zhang2005,Pechkis2013}. A resonance feature is observed near this crossover in our experiment, as illustrated in Fig.~\ref{fig1}(c) and (d). The resonance happens at about 0.14 G, evident by the longest period and the largest amplitude at this field. On the higher field side where the dynamics is dominated by $q$, the oscillations become faster with smaller amplitude. This is similar to the detuned Rabi oscillations in driven two level systems. Eventually, when the magnetic field is too large, which corresponds to the large detuning case, the oscillation amplitude becomes too small to be observed. On the lower field side, the behavior is quite different. While the amplitude also keeps decreasing, the period levels off to $\sim1/2g_1\langle n\rangle$.

As illustrated by the solid curves in Fig.~\ref{fig1}(c) and (d), these behaviors are well captured by the simulation with Eq.~(\ref{equationofmotion}). These curves are fits to the data points with the measured mean density of $\langle n\rangle\approx 3.0\times10^{13}$cm$^{-3}$ and a residual magnetization of 0.06(2) due to imperfect control of the rf pulse area in the initial state preparation. With $g_1$ as the only fitting parameter, we obtain $a_2-a_0\approx -1.00\pm0.12\ a_B$ ($a_B$ is the Bohr radius), consistent with the reported value of $a_2-a_0\approx -1.07\ a_B $ in reference~\cite{Widera2006}. The rather small but non-zero magnetization also explains the non-diverging on resonance oscillation period~\cite{Zhang2005}.

To our knowledge, the current work is the first observation of magnetically tuned spin oscillation resonance in the spin-1 $^{87}$Rb spinor gas. Previous works with BECs were performed with either an initial state with large magnetization~\cite{Chang2005} or in a quasi-one-dimensional trap ~\cite{Kronjager2005}. In the former case, the resonance does not exist~\cite{Zhang2005}. For the latter, the spin healing length $\xi = \hbar/\sqrt{2 m n |g_1|}$ is smaller than the size in the elongated direction and thus SMA is violated. Spin mixing dynamics is unstable~\cite{Zhang2005b} in this case, as perturbations can cause irreversibly conversion of the spin-dependent energy to collective excitation modes. This will lead to the formation of multiple spin domains and destroy the spin coherence within a single full oscillation, making it impossible to observe the resonance. This problem is mitigated in the current work by the tight and near 3-D crossed trap.

\subsection{Spin-2 case}

\begin{figure*}
\centering\includegraphics[width=0.65\linewidth]{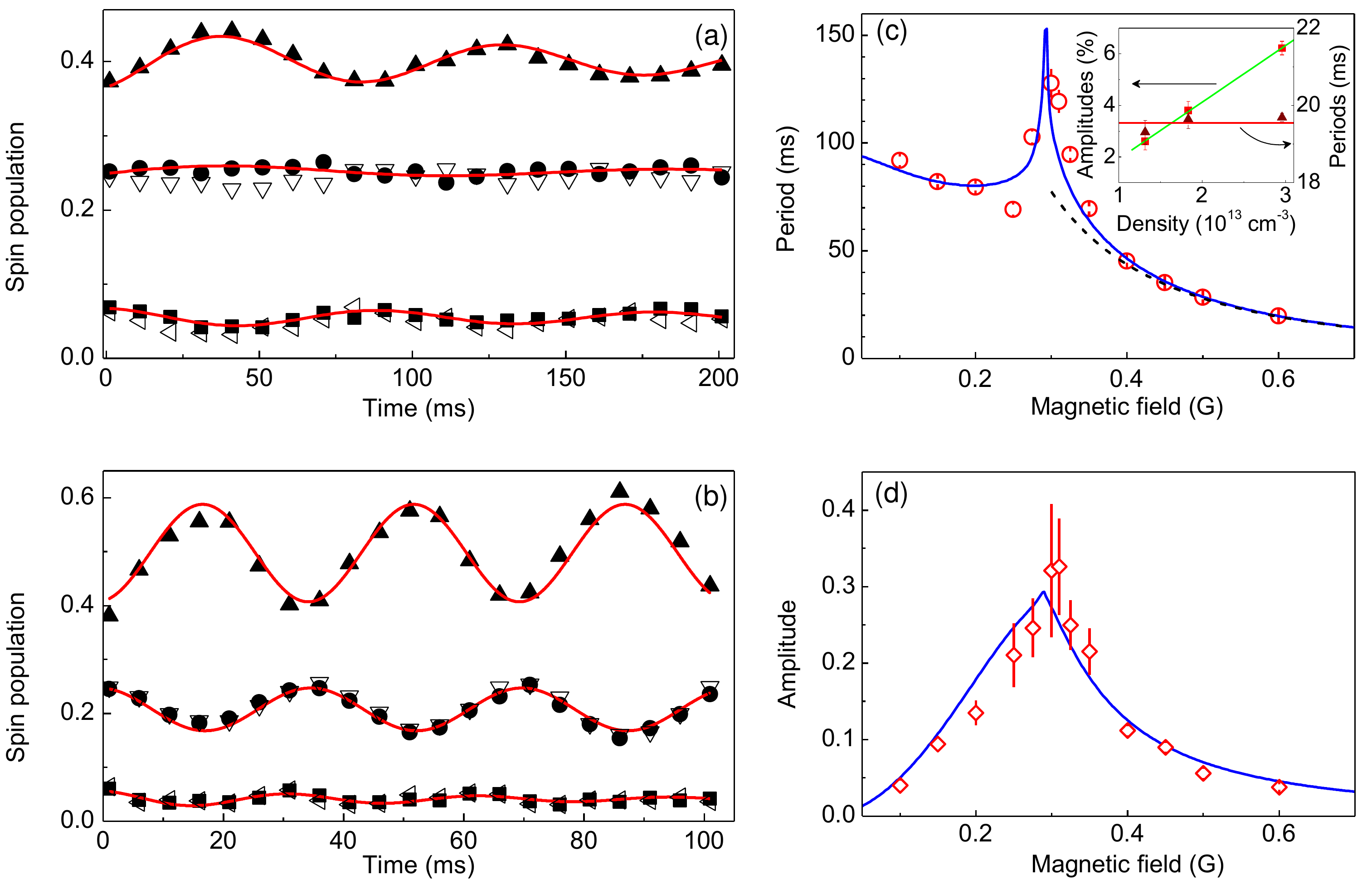}
\caption{(Color online) Spinor dynamics and its dependence on magnetic field for the $^{87}$Rb spin-2 normal gas. (a) and (b) show evolutions of $\rho_0(\blacktriangle)$, $\rho_1$($\bullet$), $\rho_{-1}$($\triangledown$), $\rho_2$($\blacksquare$), and $\rho_{-2}$($\vartriangleleft$) at 0.1 G and 0.45 G, respectively. Red solid curves in (a) and (b) are fits to the damped sinusoidal function for $\rho_0$, $\rho_1$, and $\rho_2$ only. The measured periods ($\circ$) and amplitudes ($\diamond$) of $\rho_0$ vs. magnetic fields are summarized in (c) and (d). The accompanying error bars are fitting errors. The blue solid curves here are obtained from numerical calculations with Eq.(\ref{equationofmotion}) and the black dashed curve is a plot of $\pi/q$. The inset of (c) shows the dependence of the oscillation periods (up triangle) and amplitudes (squares) on densities at 0.6 G (see text). }
\label{fig2}
\end{figure*}

The interaction between spin-2 $^{87}$Rb atoms is antiferromagnetic with $g_1>0$, $g_2<0$, and $|g_2|\ll g_1$ \cite{Schmaljohann2004}. Unlike the spin-1 case, spin-2 spinor oscillations can have more than one interaction channel and the spin-2 equation of motion has no exact analytic solutions. The spin-2 $^{87}$Rb gas is also subject to inelastic hyperfine changing collisions which greatly limit its lifetime to be about 0.5 s.

As shown in Fig.~\ref{fig2}(a) and (b), several full oscillations show up within hundreds of milliseconds. The observed behaviors are similar with those studied previously on spin-2 $^{87}$Rb BECs with the same initial state~\cite{Kronjager2006, Kronjager2008} and can be understood intuitively from the competition between $q$ and the spin-dependent interactions. Indeed, ignoring the small $g_2$ term, for the chosen initial state $\zeta_2$, approximate solutions have been obtained in references~\cite{Kronjager2006, Kronjager2008} both in the deep interaction and Zeeman regimes. These solutions can be directly generalized to our case by replacing the $g_1$ factor with $2 g_1$, as discussed in Section II.

The factor of two thermal enhancement of the spin-dependent interaction can be verified experimentally. In the Zeeman regime, spin mixing process $(0)+(0)\leftrightarrow (-1)+(1)$ dominates and the oscillation follows the fundamental period $\pi/q$~\cite{Kronjager2006,Kronjager2008}, as depicted by the dashed curve in Fig.~\ref{fig2}(c). Although the amplitude is typically small, the oscillations are quite fast so that several periods can be observed clearly. As shown in the inset of Fig.~\ref{fig2}(c), with the density varied from 1.3$\times$10$^{13}$ cm$^{-3}$ to 2.9$\times$10$^{13}$ cm$^{-3}$ at 0.6 G, the periods are nearly constant, while the amplitudes increase linearly with a slope of $0.022(2.5)/10^{13}\mathrm{cm}^{-3}$. With the best known value $a_4-a_2\approx6.95$ $a_B$~\cite{Widera2006}, this slop gives the oscillation amplitude as $2.0(3)\times3g_1\left\langle n\right\rangle/8q$. While for $F = 2$ spinor BEC, the oscillation amplitude is $3g_1\left\langle n\right\rangle/8q$ ~\cite{Kronjager2006, Kronjager2008}. This measurement thus confirms our theoretical prediction that, given a pure initial spin state, the factor of two enhancement in the $g_1$ term holds for the spin-2 thermal gas.

Fig.~\ref{fig2}(a) is taken in the interaction regime at 0.1 G, where $q$ is smaller than the $g_1$ term; while Fig.~\ref{fig2}(b) is taken at 0.45 G with $q$ much larger than the $g_1$ term. Besides the rather different oscillation amplitudes and periods, we also find that in the interaction regime, populations only oscillate between $m_F=0$ and $m_F=\pm2$ states, while those of the $m_F=\pm1$ states are nearly 0.25. In the other regime, all spin states are involved, but the $m_F=\pm2$ states have a smaller oscillation amplitude. In the interaction regime at very low field, the oscillation period is $\pi$/4$g_1\left\langle n\right\rangle$ ~\cite{Kronjager2006, Kronjager2008}, which is typically long as $g_1 \langle n \rangle$ is small. As a result, few oscillations can be observed within the sample lifetime and these oscillations also show strong damping.

We have carried out similar measurements with magnetic fields ranging from 0.1 G to 0.6 G and extracted the amplitudes and periods by fitting the oscillations to the damped sinusoidal function. As summarized in Fig.~\ref{fig2}(c) and (d), a resonance is observed near 0.3 G. Close to resonance, the approximate solution fails. Numerical calculation of the equation of motion in Eq.~(\ref{equationofmotion}) is thus necessary to fully describe the magnetic field dependence. This is performed with the $g_1$ term obtained from the enhancement factor verification above and the experimentally measured number densities. As shown by the solid curves in Fig.~\ref{fig2}(c) and (d), without any free parameters the results already agree with our measurements very well.

\section{Conclusion}
We have observed and analyzed the non-equilibrium interaction-driven collective spin mixing dynamics in ferromagnetic spin-1 and antiferromagnetic spin-2 gases of ultracold but non-condensed $^{87}$Rb atoms. These dynamics and their magnetic field dependence are proved to be the same as those found in spinor BECs under SMA, with only a factor of two enhancement in the spin-dependent interactions. In the spin-1 case, we can observe oscillations lasting for a very long time limited only by the trap lifetime. Although these oscillations already become irregular at 3 s, we nevertheless cannot tell any obvious equilibrium state is reached within 10 s. Spin domains formation are fully suppressed in both cases, but even for the spin-1 case without hyperfine changing losses, damping is still observed for most oscillations. This maybe come from the ignored collisional integral in the Boltzmann equation~\cite{Ulrich2014} as well as technical imperfections, such as the residual magnetic field gradient.

%Although in spin-1 normal gas from the fully transversely magnetized state,  the steady state at cross over region can not reach within the life time of atoms in optical dipole trap, we find that the spin-configurations approach (1/3,1/3,1/3) at low field and (1/2,1/4,1/2) at high field.

\begin{acknowledgments}

We thank Wenxian Zhang for valuable discussions. This work is supported by Hong Kong Research Grants Council (General Research Fund Projects 404712 and 403813). Z.F.X. is supported by AFOSR, ARO, DARPA OLE Program through ARO, the Charles E. Kaufman Foundation and the Pittsburgh Foundation.

\end{acknowledgments}


\begin{thebibliography}{34}%
\makeatletter
\providecommand \@ifxundefined [1]{%
 \@ifx{#1\undefined}
}%
\providecommand \@ifnum [1]{%
 \ifnum #1\expandafter \@firstoftwo
 \else \expandafter \@secondoftwo
 \fi
}%
\providecommand \@ifx [1]{%
 \ifx #1\expandafter \@firstoftwo
 \else \expandafter \@secondoftwo
 \fi
}%
\providecommand \natexlab [1]{#1}%
\providecommand \enquote  [1]{``#1''}%
\providecommand \bibnamefont  [1]{#1}%
\providecommand \bibfnamefont [1]{#1}%
\providecommand \citenamefont [1]{#1}%
\providecommand \href@noop [0]{\@secondoftwo}%
\providecommand \href [0]{\begingroup \@sanitize@url \@href}%
\providecommand \@href[1]{\@@startlink{#1}\@@href}%
\providecommand \@@href[1]{\endgroup#1\@@endlink}%
\providecommand \@sanitize@url [0]{\catcode `\\12\catcode `\$12\catcode
  `\&12\catcode `\#12\catcode `\^12\catcode `\_12\catcode `\%12\relax}%
\providecommand \@@startlink[1]{}%
\providecommand \@@endlink[0]{}%
\providecommand \url  [0]{\begingroup\@sanitize@url \@url }%
\providecommand \@url [1]{\endgroup\@href {#1}{\urlprefix }}%
\providecommand \urlprefix  [0]{URL }%
\providecommand \Eprint [0]{\href }%
\providecommand \doibase [0]{http://dx.doi.org/}%
\providecommand \selectlanguage [0]{\@gobble}%
\providecommand \bibinfo  [0]{\@secondoftwo}%
\providecommand \bibfield  [0]{\@secondoftwo}%
\providecommand \translation [1]{[#1]}%
\providecommand \BibitemOpen [0]{}%
\providecommand \bibitemStop [0]{}%
\providecommand \bibitemNoStop [0]{.\EOS\space}%
\providecommand \EOS [0]{\spacefactor3000\relax}%
\providecommand \BibitemShut  [1]{\csname bibitem#1\endcsname}%
\let\auto@bib@innerbib\@empty
%</preamble>
\bibitem [{\citenamefont {Kawaguchi}\ and\ \citenamefont
  {Ueda}(2012)}]{Kawaguchi2012}%
  \BibitemOpen
  \bibfield  {author} {\bibinfo {author} {\bibfnamefont {Y.}~\bibnamefont
  {Kawaguchi}}\ and\ \bibinfo {author} {\bibfnamefont {M.}~\bibnamefont
  {Ueda}},\ }\href@noop {} {\bibfield  {journal} {\bibinfo  {journal} {Phys.
  Rep.}\ }\textbf {\bibinfo {volume} {520}},\ \bibinfo {pages} {253 } (\bibinfo
  {year} {2012})}\BibitemShut {NoStop}%
\bibitem [{\citenamefont {Stamper-Kurn}\ and\ \citenamefont
  {Ueda}(2013)}]{Kurn2013}%
  \BibitemOpen
  \bibfield  {author} {\bibinfo {author} {\bibfnamefont {D.~M.}\ \bibnamefont
  {Stamper-Kurn}}\ and\ \bibinfo {author} {\bibfnamefont {M.}~\bibnamefont
  {Ueda}},\ }\href {\doibase 10.1103/RevModPhys.85.1191} {\bibfield  {journal}
  {\bibinfo  {journal} {Rev. Mod. Phys.}\ }\textbf {\bibinfo {volume} {85}},\
  \bibinfo {pages} {1191} (\bibinfo {year} {2013})}\BibitemShut {NoStop}%
\bibitem [{\citenamefont {Ho}(1998)}]{Ho1998}%
  \BibitemOpen
  \bibfield  {author} {\bibinfo {author} {\bibfnamefont {T.-L.}\ \bibnamefont
  {Ho}},\ }\href@noop {} {\bibfield  {journal} {\bibinfo  {journal} {Phys. Rev.
  Lett.}\ }\textbf {\bibinfo {volume} {81}},\ \bibinfo {pages} {742} (\bibinfo
  {year} {1998})}\BibitemShut {NoStop}%
\bibitem [{\citenamefont {Ohmi}\ and\ \citenamefont
  {Machida}(1998)}]{Tetsuo1998}%
  \BibitemOpen
  \bibfield  {author} {\bibinfo {author} {\bibfnamefont {T.}~\bibnamefont
  {Ohmi}}\ and\ \bibinfo {author} {\bibfnamefont {K.}~\bibnamefont {Machida}},\
  }\href@noop {} {\bibfield  {journal} {\bibinfo  {journal} {J. Phys. Soc.
  Jpn}\ }\textbf {\bibinfo {volume} {67}},\ \bibinfo {pages} {1822} (\bibinfo
  {year} {1998})}\BibitemShut {NoStop}%
\bibitem [{\citenamefont {Law}\ \emph {et~al.}(1998)\citenamefont {Law},
  \citenamefont {Pu},\ and\ \citenamefont {Bigelow}}]{Law1998}%
  \BibitemOpen
  \bibfield  {author} {\bibinfo {author} {\bibfnamefont {C.~K.}\ \bibnamefont
  {Law}}, \bibinfo {author} {\bibfnamefont {H.}~\bibnamefont {Pu}}, \ and\
  \bibinfo {author} {\bibfnamefont {N.~P.}\ \bibnamefont {Bigelow}},\
  }\href@noop {} {\bibfield  {journal} {\bibinfo  {journal} {Phys. Rev. Lett.}\
  }\textbf {\bibinfo {volume} {81}},\ \bibinfo {pages} {5257} (\bibinfo {year}
  {1998})}\BibitemShut {NoStop}%
\bibitem [{\citenamefont {Pu}\ \emph {et~al.}(1999)\citenamefont {Pu},
  \citenamefont {Law}, \citenamefont {Raghavan}, \citenamefont {Eberly},\ and\
  \citenamefont {Bigelow}}]{Pu1999}%
  \BibitemOpen
  \bibfield  {author} {\bibinfo {author} {\bibfnamefont {H.}~\bibnamefont
  {Pu}}, \bibinfo {author} {\bibfnamefont {C.~K.}\ \bibnamefont {Law}},
  \bibinfo {author} {\bibfnamefont {S.}~\bibnamefont {Raghavan}}, \bibinfo
  {author} {\bibfnamefont {J.~H.}\ \bibnamefont {Eberly}}, \ and\ \bibinfo
  {author} {\bibfnamefont {N.~P.}\ \bibnamefont {Bigelow}},\ }\href {\doibase
  10.1103/PhysRevA.60.1463} {\bibfield  {journal} {\bibinfo  {journal} {Phys.
  Rev. A}\ }\textbf {\bibinfo {volume} {60}},\ \bibinfo {pages} {1463}
  (\bibinfo {year} {1999})}\BibitemShut {NoStop}%
\bibitem [{\citenamefont {Zhang}\ \emph
  {et~al.}(2005{\natexlab{a}})\citenamefont {Zhang}, \citenamefont {Zhou},
  \citenamefont {Chang}, \citenamefont {Chapman},\ and\ \citenamefont
  {You}}]{Zhang2005}%
  \BibitemOpen
  \bibfield  {author} {\bibinfo {author} {\bibfnamefont {W.}~\bibnamefont
  {Zhang}}, \bibinfo {author} {\bibfnamefont {D.~L.}\ \bibnamefont {Zhou}},
  \bibinfo {author} {\bibfnamefont {M.-S.}\ \bibnamefont {Chang}}, \bibinfo
  {author} {\bibfnamefont {M.~S.}\ \bibnamefont {Chapman}}, \ and\ \bibinfo
  {author} {\bibfnamefont {L.}~\bibnamefont {You}},\ }\href {\doibase
  10.1103/PhysRevA.72.013602} {\bibfield  {journal} {\bibinfo  {journal} {Phys.
  Rev. A}\ }\textbf {\bibinfo {volume} {72}},\ \bibinfo {pages} {013602}
  (\bibinfo {year} {2005}{\natexlab{a}})}\BibitemShut {NoStop}%
\bibitem [{\citenamefont {Kronj\"ager}\ \emph {et~al.}(2005)\citenamefont
  {Kronj\"ager}, \citenamefont {Becker}, \citenamefont {Brinkmann},
  \citenamefont {Walser}, \citenamefont {Navez}, \citenamefont {Bongs},\ and\
  \citenamefont {Sengstock}}]{Kronjager2005}%
  \BibitemOpen
  \bibfield  {author} {\bibinfo {author} {\bibfnamefont {J.}~\bibnamefont
  {Kronj\"ager}}, \bibinfo {author} {\bibfnamefont {C.}~\bibnamefont {Becker}},
  \bibinfo {author} {\bibfnamefont {M.}~\bibnamefont {Brinkmann}}, \bibinfo
  {author} {\bibfnamefont {R.}~\bibnamefont {Walser}}, \bibinfo {author}
  {\bibfnamefont {P.}~\bibnamefont {Navez}}, \bibinfo {author} {\bibfnamefont
  {K.}~\bibnamefont {Bongs}}, \ and\ \bibinfo {author} {\bibfnamefont
  {K.}~\bibnamefont {Sengstock}},\ }\href {\doibase 10.1103/PhysRevA.72.063619}
  {\bibfield  {journal} {\bibinfo  {journal} {Phys. Rev. A}\ }\textbf {\bibinfo
  {volume} {72}},\ \bibinfo {pages} {063619} (\bibinfo {year}
  {2005})}\BibitemShut {NoStop}%
\bibitem [{\citenamefont {Chang}\ \emph {et~al.}(2005)\citenamefont {Chang},
  \citenamefont {Qin}, \citenamefont {Zhang}, \citenamefont {You},\ and\
  \citenamefont {Chapman}}]{Chang2005}%
  \BibitemOpen
  \bibfield  {author} {\bibinfo {author} {\bibfnamefont {M.-S.}\ \bibnamefont
  {Chang}}, \bibinfo {author} {\bibfnamefont {Q.}~\bibnamefont {Qin}}, \bibinfo
  {author} {\bibfnamefont {W.}~\bibnamefont {Zhang}}, \bibinfo {author}
  {\bibfnamefont {L.}~\bibnamefont {You}}, \ and\ \bibinfo {author}
  {\bibfnamefont {M.~S.}\ \bibnamefont {Chapman}},\ }\href@noop {} {\bibfield
  {journal} {\bibinfo  {journal} {Nat. Phys.}\ }\textbf {\bibinfo {volume}
  {1}},\ \bibinfo {pages} {111} (\bibinfo {year} {2005})}\BibitemShut {NoStop}%
\bibitem [{\citenamefont {Widera}\ \emph {et~al.}(2005)\citenamefont {Widera},
  \citenamefont {Gerbier}, \citenamefont {F\"olling}, \citenamefont {Gericke},
  \citenamefont {Mandel},\ and\ \citenamefont {Bloch}}]{Widera2005}%
  \BibitemOpen
  \bibfield  {author} {\bibinfo {author} {\bibfnamefont {A.}~\bibnamefont
  {Widera}}, \bibinfo {author} {\bibfnamefont {F.}~\bibnamefont {Gerbier}},
  \bibinfo {author} {\bibfnamefont {S.}~\bibnamefont {F\"olling}}, \bibinfo
  {author} {\bibfnamefont {T.}~\bibnamefont {Gericke}}, \bibinfo {author}
  {\bibfnamefont {O.}~\bibnamefont {Mandel}}, \ and\ \bibinfo {author}
  {\bibfnamefont {I.}~\bibnamefont {Bloch}},\ }\href {\doibase
  10.1103/PhysRevLett.95.190405} {\bibfield  {journal} {\bibinfo  {journal}
  {Phys. Rev. Lett.}\ }\textbf {\bibinfo {volume} {95}},\ \bibinfo {pages}
  {190405} (\bibinfo {year} {2005})}\BibitemShut {NoStop}%
\bibitem [{\citenamefont {Kronjager}\ \emph {et~al.}(2006)\citenamefont
  {Kronjager}, \citenamefont {Becker}, \citenamefont {Navez}, \citenamefont
  {Bongs},\ and\ \citenamefont {Sengstock}}]{Kronjager2006}%
  \BibitemOpen
  \bibfield  {author} {\bibinfo {author} {\bibfnamefont {J.}~\bibnamefont
  {Kronjager}}, \bibinfo {author} {\bibfnamefont {C.}~\bibnamefont {Becker}},
  \bibinfo {author} {\bibfnamefont {P.}~\bibnamefont {Navez}}, \bibinfo
  {author} {\bibfnamefont {K.}~\bibnamefont {Bongs}}, \ and\ \bibinfo {author}
  {\bibfnamefont {K.}~\bibnamefont {Sengstock}},\ }\href {\doibase
  10.1103/PhysRevLett.97.110404} {\bibfield  {journal} {\bibinfo  {journal}
  {Phys. Rev. Lett.}\ }\textbf {\bibinfo {volume} {97}},\ \bibinfo {pages}
  {110404} (\bibinfo {year} {2006})}\BibitemShut {NoStop}%
\bibitem [{\citenamefont {Black}\ \emph {et~al.}(2007)\citenamefont {Black},
  \citenamefont {Gomez}, \citenamefont {Turner}, \citenamefont {Jung},\ and\
  \citenamefont {Lett}}]{Black2007}%
  \BibitemOpen
  \bibfield  {author} {\bibinfo {author} {\bibfnamefont {A.~T.}\ \bibnamefont
  {Black}}, \bibinfo {author} {\bibfnamefont {E.}~\bibnamefont {Gomez}},
  \bibinfo {author} {\bibfnamefont {L.~D.}\ \bibnamefont {Turner}}, \bibinfo
  {author} {\bibfnamefont {S.}~\bibnamefont {Jung}}, \ and\ \bibinfo {author}
  {\bibfnamefont {P.~D.}\ \bibnamefont {Lett}},\ }\href@noop {} {\bibfield
  {journal} {\bibinfo  {journal} {Phys. Rev. Lett.}\ }\textbf {\bibinfo
  {volume} {99}},\ \bibinfo {pages} {070403} (\bibinfo {year}
  {2007})}\BibitemShut {NoStop}%
\bibitem [{\citenamefont {Deutsch}\ \emph {et~al.}(2010)\citenamefont
  {Deutsch}, \citenamefont {Ramirez-Martinez}, \citenamefont {Lacro\^ute},
  \citenamefont {Reinhard}, \citenamefont {Schneider}, \citenamefont {Fuchs},
  \citenamefont {Pi\'echon}, \citenamefont {Lalo\"e}, \citenamefont {Reichel},\
  and\ \citenamefont {Rosenbusch}}]{Deutsch2010}%
  \BibitemOpen
  \bibfield  {author} {\bibinfo {author} {\bibfnamefont {C.}~\bibnamefont
  {Deutsch}}, \bibinfo {author} {\bibfnamefont {F.}~\bibnamefont
  {Ramirez-Martinez}}, \bibinfo {author} {\bibfnamefont {C.}~\bibnamefont
  {Lacro\^ute}}, \bibinfo {author} {\bibfnamefont {F.}~\bibnamefont
  {Reinhard}}, \bibinfo {author} {\bibfnamefont {T.}~\bibnamefont {Schneider}},
  \bibinfo {author} {\bibfnamefont {J.~N.}\ \bibnamefont {Fuchs}}, \bibinfo
  {author} {\bibfnamefont {F.}~\bibnamefont {Pi\'echon}}, \bibinfo {author}
  {\bibfnamefont {F.}~\bibnamefont {Lalo\"e}}, \bibinfo {author} {\bibfnamefont
  {J.}~\bibnamefont {Reichel}}, \ and\ \bibinfo {author} {\bibfnamefont
  {P.}~\bibnamefont {Rosenbusch}},\ }\href@noop {} {\bibfield  {journal}
  {\bibinfo  {journal} {Phys. Rev. Lett.}\ }\textbf {\bibinfo {volume} {105}},\
  \bibinfo {pages} {020401} (\bibinfo {year} {2010})}\BibitemShut {NoStop}%
\bibitem [{\citenamefont {Kleine~B\"uning}\ \emph {et~al.}(2011)\citenamefont
  {Kleine~B\"uning}, \citenamefont {Will}, \citenamefont {Ertmer},
  \citenamefont {Rasel}, \citenamefont {Arlt}, \citenamefont {Klempt},
  \citenamefont {Ramirez-Martinez}, \citenamefont {Pi\'echon},\ and\
  \citenamefont {Rosenbusch}}]{Buning2011}%
  \BibitemOpen
  \bibfield  {author} {\bibinfo {author} {\bibfnamefont {G.}~\bibnamefont
  {Kleine~B\"uning}}, \bibinfo {author} {\bibfnamefont {J.}~\bibnamefont
  {Will}}, \bibinfo {author} {\bibfnamefont {W.}~\bibnamefont {Ertmer}},
  \bibinfo {author} {\bibfnamefont {E.}~\bibnamefont {Rasel}}, \bibinfo
  {author} {\bibfnamefont {J.}~\bibnamefont {Arlt}}, \bibinfo {author}
  {\bibfnamefont {C.}~\bibnamefont {Klempt}}, \bibinfo {author} {\bibfnamefont
  {F.}~\bibnamefont {Ramirez-Martinez}}, \bibinfo {author} {\bibfnamefont
  {F.}~\bibnamefont {Pi\'echon}}, \ and\ \bibinfo {author} {\bibfnamefont
  {P.}~\bibnamefont {Rosenbusch}},\ }\href {\doibase
  10.1103/PhysRevLett.106.240801} {\bibfield  {journal} {\bibinfo  {journal}
  {Phys. Rev. Lett.}\ }\textbf {\bibinfo {volume} {106}},\ \bibinfo {pages}
  {240801} (\bibinfo {year} {2011})}\BibitemShut {NoStop}%
\bibitem [{\citenamefont {Pechkis}\ \emph {et~al.}(2013)\citenamefont
  {Pechkis}, \citenamefont {Wrubel}, \citenamefont {Schwettmann}, \citenamefont
  {Griffin}, \citenamefont {Barnett}, \citenamefont {Tiesinga},\ and\
  \citenamefont {Lett}}]{Pechkis2013}%
  \BibitemOpen
  \bibfield  {author} {\bibinfo {author} {\bibfnamefont {H.~K.}\ \bibnamefont
  {Pechkis}}, \bibinfo {author} {\bibfnamefont {J.~P.}\ \bibnamefont {Wrubel}},
  \bibinfo {author} {\bibfnamefont {A.}~\bibnamefont {Schwettmann}}, \bibinfo
  {author} {\bibfnamefont {P.~F.}\ \bibnamefont {Griffin}}, \bibinfo {author}
  {\bibfnamefont {R.}~\bibnamefont {Barnett}}, \bibinfo {author} {\bibfnamefont
  {E.}~\bibnamefont {Tiesinga}}, \ and\ \bibinfo {author} {\bibfnamefont
  {P.~D.}\ \bibnamefont {Lett}},\ }\href {\doibase
  10.1103/PhysRevLett.111.025301} {\bibfield  {journal} {\bibinfo  {journal}
  {Phys. Rev. Lett.}\ }\textbf {\bibinfo {volume} {111}},\ \bibinfo {pages}
  {025301} (\bibinfo {year} {2013})}\BibitemShut {NoStop}%
\bibitem [{\citenamefont {Krauser}\ \emph {et~al.}(2014)\citenamefont
  {Krauser}, \citenamefont {Ebling}, \citenamefont {Fl?schner}, \citenamefont
  {Heinze}, \citenamefont {Sengstock}, \citenamefont {Lewenstein},
  \citenamefont {Eckardt},\ and\ \citenamefont {Becker}}]{Krauser2014}%
  \BibitemOpen
  \bibfield  {author} {\bibinfo {author} {\bibfnamefont {J.~S.}\ \bibnamefont
  {Krauser}}, \bibinfo {author} {\bibfnamefont {U.}~\bibnamefont {Ebling}},
  \bibinfo {author} {\bibfnamefont {N.}~\bibnamefont {Fl?schner}}, \bibinfo
  {author} {\bibfnamefont {J.}~\bibnamefont {Heinze}}, \bibinfo {author}
  {\bibfnamefont {K.}~\bibnamefont {Sengstock}}, \bibinfo {author}
  {\bibfnamefont {M.}~\bibnamefont {Lewenstein}}, \bibinfo {author}
  {\bibfnamefont {A.}~\bibnamefont {Eckardt}}, \ and\ \bibinfo {author}
  {\bibfnamefont {C.}~\bibnamefont {Becker}},\ }\href {\doibase
  10.1126/science.1244059} {\bibfield  {journal} {\bibinfo  {journal}
  {Science}\ }\textbf {\bibinfo {volume} {343}},\ \bibinfo {pages} {157}
  (\bibinfo {year} {2014})}\BibitemShut {NoStop}%
\bibitem [{\citenamefont {Ebling}\ \emph {et~al.}(2014)\citenamefont {Ebling},
  \citenamefont {Krauser}, \citenamefont {Fl\"aschner}, \citenamefont
  {Sengstock}, \citenamefont {Becker}, \citenamefont {Lewenstein},\ and\
  \citenamefont {Eckardt}}]{Ulrich2014}%
  \BibitemOpen
  \bibfield  {author} {\bibinfo {author} {\bibfnamefont {U.}~\bibnamefont
  {Ebling}}, \bibinfo {author} {\bibfnamefont {J.~S.}\ \bibnamefont {Krauser}},
  \bibinfo {author} {\bibfnamefont {N.}~\bibnamefont {Fl\"aschner}}, \bibinfo
  {author} {\bibfnamefont {K.}~\bibnamefont {Sengstock}}, \bibinfo {author}
  {\bibfnamefont {C.}~\bibnamefont {Becker}}, \bibinfo {author} {\bibfnamefont
  {M.}~\bibnamefont {Lewenstein}}, \ and\ \bibinfo {author} {\bibfnamefont
  {A.}~\bibnamefont {Eckardt}},\ }\href {\doibase 10.1103/PhysRevX.4.021011}
  {\bibfield  {journal} {\bibinfo  {journal} {Phys. Rev. X}\ }\textbf {\bibinfo
  {volume} {4}},\ \bibinfo {pages} {021011} (\bibinfo {year}
  {2014})}\BibitemShut {NoStop}%
\bibitem [{\citenamefont {Stenger}\ \emph {et~al.}(1998)\citenamefont
  {Stenger}, \citenamefont {Inouye}, \citenamefont {Stamper-Kurn},
  \citenamefont {Miesner}, \citenamefont {Chikkatur},\ and\ \citenamefont
  {Ketterle}}]{Stenger1998}%
  \BibitemOpen
  \bibfield  {author} {\bibinfo {author} {\bibfnamefont {J.}~\bibnamefont
  {Stenger}}, \bibinfo {author} {\bibfnamefont {S.}~\bibnamefont {Inouye}},
  \bibinfo {author} {\bibfnamefont {D.~M.}\ \bibnamefont {Stamper-Kurn}},
  \bibinfo {author} {\bibfnamefont {H.-J.}\ \bibnamefont {Miesner}}, \bibinfo
  {author} {\bibfnamefont {A.~P.}\ \bibnamefont {Chikkatur}}, \ and\ \bibinfo
  {author} {\bibfnamefont {W.}~\bibnamefont {Ketterle}},\ }\href@noop {}
  {\bibfield  {journal} {\bibinfo  {journal} {Nature}\ }\textbf {\bibinfo
  {volume} {396}},\ \bibinfo {pages} {345} (\bibinfo {year}
  {1998})}\BibitemShut {NoStop}%
\bibitem [{\citenamefont {Klausen}\ \emph {et~al.}(2001)\citenamefont
  {Klausen}, \citenamefont {Bohn},\ and\ \citenamefont {Greene}}]{Klausen2001}%
  \BibitemOpen
  \bibfield  {author} {\bibinfo {author} {\bibfnamefont {N.~N.}\ \bibnamefont
  {Klausen}}, \bibinfo {author} {\bibfnamefont {J.~L.}\ \bibnamefont {Bohn}}, \
  and\ \bibinfo {author} {\bibfnamefont {C.~H.}\ \bibnamefont {Greene}},\
  }\href@noop {} {\bibfield  {journal} {\bibinfo  {journal} {Phys. Rev. A}\
  }\textbf {\bibinfo {volume} {64}},\ \bibinfo {pages} {053602} (\bibinfo
  {year} {2001})}\BibitemShut {NoStop}%
\bibitem [{\citenamefont {van Kempen}\ \emph {et~al.}(2002)\citenamefont {van
  Kempen}, \citenamefont {Kokkelmans}, \citenamefont {Heinzen},\ and\
  \citenamefont {Verhaar}}]{Kempen2002}%
  \BibitemOpen
  \bibfield  {author} {\bibinfo {author} {\bibfnamefont {E.~G.~M.}\
  \bibnamefont {van Kempen}}, \bibinfo {author} {\bibfnamefont {S.~J. J.
  M.~F.}\ \bibnamefont {Kokkelmans}}, \bibinfo {author} {\bibfnamefont {D.~J.}\
  \bibnamefont {Heinzen}}, \ and\ \bibinfo {author} {\bibfnamefont {B.~J.}\
  \bibnamefont {Verhaar}},\ }\href@noop {} {\bibfield  {journal} {\bibinfo
  {journal} {Phys. Rev. Lett.}\ }\textbf {\bibinfo {volume} {88}},\ \bibinfo
  {pages} {093201} (\bibinfo {year} {2002})}\BibitemShut {NoStop}%
\bibitem [{\citenamefont {Koashi}\ and\ \citenamefont
  {Ueda}(2000)}]{Koashi2000}%
  \BibitemOpen
  \bibfield  {author} {\bibinfo {author} {\bibfnamefont {M.}~\bibnamefont
  {Koashi}}\ and\ \bibinfo {author} {\bibfnamefont {M.}~\bibnamefont {Ueda}},\
  }\href@noop {} {\bibfield  {journal} {\bibinfo  {journal} {Phys. Rev. Lett.}\
  }\textbf {\bibinfo {volume} {84}},\ \bibinfo {pages} {1066} (\bibinfo {year}
  {2000})}\BibitemShut {NoStop}%
\bibitem [{\citenamefont {Ciobanu}\ \emph {et~al.}(2000)\citenamefont
  {Ciobanu}, \citenamefont {Yip},\ and\ \citenamefont {Ho}}]{Ciobanu2000}%
  \BibitemOpen
  \bibfield  {author} {\bibinfo {author} {\bibfnamefont {C.~V.}\ \bibnamefont
  {Ciobanu}}, \bibinfo {author} {\bibfnamefont {S.-K.}\ \bibnamefont {Yip}}, \
  and\ \bibinfo {author} {\bibfnamefont {T.-L.}\ \bibnamefont {Ho}},\ }\href
  {\doibase 10.1103/PhysRevA.61.033607} {\bibfield  {journal} {\bibinfo
  {journal} {Phys. Rev. A}\ }\textbf {\bibinfo {volume} {61}},\ \bibinfo
  {pages} {033607} (\bibinfo {year} {2000})}\BibitemShut {NoStop}%
\bibitem [{\citenamefont {Endo}\ and\ \citenamefont {Nikuni}(2008)}]{Endo2008}%
  \BibitemOpen
  \bibfield  {author} {\bibinfo {author} {\bibfnamefont {Y.}~\bibnamefont
  {Endo}}\ and\ \bibinfo {author} {\bibfnamefont {T.}~\bibnamefont {Nikuni}},\
  }\href@noop {} {\bibfield  {journal} {\bibinfo  {journal} {J. Low Temp.
  Phys.}\ }\textbf {\bibinfo {volume} {152}},\ \bibinfo {pages} {21} (\bibinfo
  {year} {2008})}\BibitemShut {NoStop}%
\bibitem [{\citenamefont {Natu}\ and\ \citenamefont
  {Mueller}(2010)}]{Natu2010}%
  \BibitemOpen
  \bibfield  {author} {\bibinfo {author} {\bibfnamefont {S.~S.}\ \bibnamefont
  {Natu}}\ and\ \bibinfo {author} {\bibfnamefont {E.~J.}\ \bibnamefont
  {Mueller}},\ }\href {\doibase 10.1103/PhysRevA.81.053617} {\bibfield
  {journal} {\bibinfo  {journal} {Phys. Rev. A}\ }\textbf {\bibinfo {volume}
  {81}},\ \bibinfo {pages} {053617} (\bibinfo {year} {2010})}\BibitemShut
  {NoStop}%
\bibitem [{\citenamefont {Schmaljohann}\ \emph {et~al.}(2004)\citenamefont
  {Schmaljohann}, \citenamefont {Erhard}, \citenamefont {Kronj\"ager},
  \citenamefont {Kottke}, \citenamefont {van Staa}, \citenamefont
  {Cacciapuoti}, \citenamefont {Arlt}, \citenamefont {Bongs},\ and\
  \citenamefont {Sengstock}}]{Schmaljohann2004}%
  \BibitemOpen
  \bibfield  {author} {\bibinfo {author} {\bibfnamefont {H.}~\bibnamefont
  {Schmaljohann}}, \bibinfo {author} {\bibfnamefont {M.}~\bibnamefont
  {Erhard}}, \bibinfo {author} {\bibfnamefont {J.}~\bibnamefont {Kronj\"ager}},
  \bibinfo {author} {\bibfnamefont {M.}~\bibnamefont {Kottke}}, \bibinfo
  {author} {\bibfnamefont {S.}~\bibnamefont {van Staa}}, \bibinfo {author}
  {\bibfnamefont {L.}~\bibnamefont {Cacciapuoti}}, \bibinfo {author}
  {\bibfnamefont {J.~J.}\ \bibnamefont {Arlt}}, \bibinfo {author}
  {\bibfnamefont {K.}~\bibnamefont {Bongs}}, \ and\ \bibinfo {author}
  {\bibfnamefont {K.}~\bibnamefont {Sengstock}},\ }\href {\doibase
  10.1103/PhysRevLett.92.040402} {\bibfield  {journal} {\bibinfo  {journal}
  {Phys. Rev. Lett.}\ }\textbf {\bibinfo {volume} {92}},\ \bibinfo {pages}
  {040402} (\bibinfo {year} {2004})}\BibitemShut {NoStop}%
\bibitem [{\citenamefont {Xiong}\ \emph
  {et~al.}(2013{\natexlab{a}})\citenamefont {Xiong}, \citenamefont {Wang},
  \citenamefont {Li}, \citenamefont {Lam},\ and\ \citenamefont
  {Wang}}]{Xiong2013a}%
  \BibitemOpen
  \bibfield  {author} {\bibinfo {author} {\bibfnamefont {D.}~\bibnamefont
  {Xiong}}, \bibinfo {author} {\bibfnamefont {F.}~\bibnamefont {Wang}},
  \bibinfo {author} {\bibfnamefont {X.}~\bibnamefont {Li}}, \bibinfo {author}
  {\bibfnamefont {T.-F.}\ \bibnamefont {Lam}}, \ and\ \bibinfo {author}
  {\bibfnamefont {D.}~\bibnamefont {Wang}},\ }\href@noop {} {\bibfield
  {journal} {\bibinfo  {journal} {arXiv:1303.0333 [cond-mat.quant-gas]}\ }
  (\bibinfo {year} {2013}{\natexlab{a}})}\BibitemShut {NoStop}%
\bibitem [{\citenamefont {Wang}\ \emph {et~al.}(2013)\citenamefont {Wang},
  \citenamefont {Xiong}, \citenamefont {Li}, \citenamefont {Wang},\ and\
  \citenamefont {Tiemann}}]{Wang2013}%
  \BibitemOpen
  \bibfield  {author} {\bibinfo {author} {\bibfnamefont {F.}~\bibnamefont
  {Wang}}, \bibinfo {author} {\bibfnamefont {D.}~\bibnamefont {Xiong}},
  \bibinfo {author} {\bibfnamefont {X.}~\bibnamefont {Li}}, \bibinfo {author}
  {\bibfnamefont {D.}~\bibnamefont {Wang}}, \ and\ \bibinfo {author}
  {\bibfnamefont {E.}~\bibnamefont {Tiemann}},\ }\href@noop {} {\bibfield
  {journal} {\bibinfo  {journal} {Phys. Rev. A}\ }\textbf {\bibinfo {volume}
  {87}},\ \bibinfo {pages} {050702} (\bibinfo {year} {2013})}\BibitemShut
  {NoStop}%
\bibitem [{\citenamefont {Lin}\ \emph {et~al.}(2009)\citenamefont {Lin},
  \citenamefont {Perry}, \citenamefont {Compton}, \citenamefont {Spielman},\
  and\ \citenamefont {Porto}}]{Lin2009}%
  \BibitemOpen
  \bibfield  {author} {\bibinfo {author} {\bibfnamefont {Y.-J.}\ \bibnamefont
  {Lin}}, \bibinfo {author} {\bibfnamefont {A.~R.}\ \bibnamefont {Perry}},
  \bibinfo {author} {\bibfnamefont {R.~L.}\ \bibnamefont {Compton}}, \bibinfo
  {author} {\bibfnamefont {I.~B.}\ \bibnamefont {Spielman}}, \ and\ \bibinfo
  {author} {\bibfnamefont {J.~V.}\ \bibnamefont {Porto}},\ }\href@noop {}
  {\bibfield  {journal} {\bibinfo  {journal} {Phys. Rev. A}\ }\textbf {\bibinfo
  {volume} {79}},\ \bibinfo {pages} {063631} (\bibinfo {year}
  {2009})}\BibitemShut {NoStop}%
\bibitem [{\citenamefont {Xiong}\ \emph
  {et~al.}(2013{\natexlab{b}})\citenamefont {Xiong}, \citenamefont {Li},
  \citenamefont {Wang},\ and\ \citenamefont {Wang}}]{Xiong2013b}%
  \BibitemOpen
  \bibfield  {author} {\bibinfo {author} {\bibfnamefont {D.}~\bibnamefont
  {Xiong}}, \bibinfo {author} {\bibfnamefont {X.}~\bibnamefont {Li}}, \bibinfo
  {author} {\bibfnamefont {F.}~\bibnamefont {Wang}}, \ and\ \bibinfo {author}
  {\bibfnamefont {D.}~\bibnamefont {Wang}},\ }\href@noop {} {\bibfield
  {journal} {\bibinfo  {journal} {arXiv:1305.7091}\ }
  (\bibinfo {year} {2013}{\natexlab{b}})}\BibitemShut {NoStop}%
\bibitem [{\citenamefont {Reinaudi}\ \emph {et~al.}(2007)\citenamefont
  {Reinaudi}, \citenamefont {Lahaye}, \citenamefont {Wang},\ and\ \citenamefont
  {Gu\'{e}ry-Odelin}}]{Reinaudi2007}%
  \BibitemOpen
  \bibfield  {author} {\bibinfo {author} {\bibfnamefont {G.}~\bibnamefont
  {Reinaudi}}, \bibinfo {author} {\bibfnamefont {T.}~\bibnamefont {Lahaye}},
  \bibinfo {author} {\bibfnamefont {Z.}~\bibnamefont {Wang}}, \ and\ \bibinfo
  {author} {\bibfnamefont {D.}~\bibnamefont {Gu\'{e}ry-Odelin}},\ }\href
  {\doibase 10.1364/OL.32.003143} {\bibfield  {journal} {\bibinfo  {journal}
  {Opt. Lett.}\ }\textbf {\bibinfo {volume} {32}},\ \bibinfo {pages} {3143}
  (\bibinfo {year} {2007})}\BibitemShut {NoStop}%
\bibitem [{\citenamefont {Kwon}\ \emph {et~al.}(2012)\citenamefont {Kwon},
  \citenamefont {Choi},\ and\ \citenamefont {Shin}}]{Kwon2012}%
  \BibitemOpen
  \bibfield  {author} {\bibinfo {author} {\bibfnamefont {W.}~\bibnamefont
  {Kwon}}, \bibinfo {author} {\bibfnamefont {J.-y.}\ \bibnamefont {Choi}}, \
  and\ \bibinfo {author} {\bibfnamefont {Y.-i.}\ \bibnamefont {Shin}},\
  }\href@noop {} {\bibfield  {journal} {\bibinfo  {journal} {J. Kore. Phys.
  Soc.}\ }\textbf {\bibinfo {volume} {61}},\ \bibinfo {pages} {1970} (\bibinfo
  {year} {2012})}\BibitemShut {NoStop}%
\bibitem [{\citenamefont {Widera}\ \emph {et~al.}(2006)\citenamefont {Widera},
  \citenamefont {Gerbier}, \citenamefont {Flling}, \citenamefont {Gericke},
  \citenamefont {Mandel},\ and\ \citenamefont {Bloch}}]{Widera2006}%
  \BibitemOpen
  \bibfield  {author} {\bibinfo {author} {\bibfnamefont {A.}~\bibnamefont
  {Widera}}, \bibinfo {author} {\bibfnamefont {F.}~\bibnamefont {Gerbier}},
  \bibinfo {author} {\bibfnamefont {S.}~\bibnamefont {Flling}}, \bibinfo
  {author} {\bibfnamefont {T.}~\bibnamefont {Gericke}}, \bibinfo {author}
  {\bibfnamefont {O.}~\bibnamefont {Mandel}}, \ and\ \bibinfo {author}
  {\bibfnamefont {I.}~\bibnamefont {Bloch}},\ }\href@noop {} {\bibfield
  {journal} {\bibinfo  {journal} {New J. Phys.}\ }\textbf {\bibinfo {volume}
  {8}},\ \bibinfo {pages} {152} (\bibinfo {year} {2006})}\BibitemShut {NoStop}%
\bibitem [{\citenamefont {Zhang}\ \emph
  {et~al.}(2005{\natexlab{b}})\citenamefont {Zhang}, \citenamefont {Zhou},
  \citenamefont {Chang}, \citenamefont {Chapman},\ and\ \citenamefont
  {You}}]{Zhang2005b}%
  \BibitemOpen
  \bibfield  {author} {\bibinfo {author} {\bibfnamefont {W.}~\bibnamefont
  {Zhang}}, \bibinfo {author} {\bibfnamefont {D.~L.}\ \bibnamefont {Zhou}},
  \bibinfo {author} {\bibfnamefont {M.-S.}\ \bibnamefont {Chang}}, \bibinfo
  {author} {\bibfnamefont {M.~S.}\ \bibnamefont {Chapman}}, \ and\ \bibinfo
  {author} {\bibfnamefont {L.}~\bibnamefont {You}},\ }\href {\doibase
  10.1103/PhysRevLett.95.180403} {\bibfield  {journal} {\bibinfo  {journal}
  {Phys. Rev. Lett.}\ }\textbf {\bibinfo {volume} {95}},\ \bibinfo {pages}
  {180403} (\bibinfo {year} {2005}{\natexlab{b}})}\BibitemShut {NoStop}%
\bibitem [{\citenamefont {Kronj\"ager}\ \emph {et~al.}(2008)\citenamefont
  {Kronj\"ager}, \citenamefont {Becker}, \citenamefont {Navez}, \citenamefont
  {Bongs},\ and\ \citenamefont {Sengstock}}]{Kronjager2008}%
  \BibitemOpen
  \bibfield  {author} {\bibinfo {author} {\bibfnamefont {J.}~\bibnamefont
  {Kronj\"ager}}, \bibinfo {author} {\bibfnamefont {C.}~\bibnamefont {Becker}},
  \bibinfo {author} {\bibfnamefont {P.}~\bibnamefont {Navez}}, \bibinfo
  {author} {\bibfnamefont {K.}~\bibnamefont {Bongs}}, \ and\ \bibinfo {author}
  {\bibfnamefont {K.}~\bibnamefont {Sengstock}},\ }\href {\doibase
  10.1103/PhysRevLett.100.189901} {\bibfield  {journal} {\bibinfo  {journal}
  {Phys. Rev. Lett.}\ }\textbf {\bibinfo {volume} {100}},\ \bibinfo {pages}
  {189901} (\bibinfo {year} {2008})}\BibitemShut {NoStop}%
\end{thebibliography}
\end{document}